\documentclass[pre,aps,twocolumn]{revtex4}
\begin{document}
\title{An exact sampling scheme for Brownian motion in the presence of a magnetic field}
\author{Mini P. Balakrishnan$^1$, M. C. Valsakumar$^2$ and P. Rameshan$^1$}
\affiliation{$^1$ Department of Physics, University of Calicut, Calicut University PO, Pin 673635, India\\
$^2$ Materials Science Division, IGCAR, Kalpakkam, Pin 603102, India}
\begin{abstract}
Langevin equation pertinent  to  diffusion limited aggregation  of charged particles in the
presence of an external magnetic field is solved exactly. The solution involves correlated
random variables. A new scheme for exactly  sampling the components of the position and velocity is
proposed.
\end{abstract}
\date{\today}
\maketitle

\section{Introduction}

The diffusion limited aggregation(DLA) model \cite{witten} and  some of its variants
have been used to model pattern formation in diverse contexts such as electrochemical deposition,
dissolution and erosion of porous media, gelation, fracture,  dielectric breakdown etc
\cite{bunde,xi,ausloos,meakin,nietmeyer}. These patterns, in general, have a
fractal structure and the study of  DLA clusters have resulted in the understanding
of various aspects of fractals such as fractal dimension, multifractality and the scaling laws.
The insight obtained  from the analysis of DLA clusters has led to the application of
this model to the study of the scaling behavior of two-dimensional quantum gravity.
Yet a complete theoretical understanding of DLA has not been achieved to date. Recently,
a few groups \cite{matsushita,sawada}
 have carried out experimental investigation of  the influence of an external magnetic field on DLA of
  charged particles. They observed a bending of the branches of the DLA cluster, which could be attributed to the
  Lorentz force, as well as a conspicuous change in the morphology of the clusters with increase
   in the strength of the magnetic field. As a prelude to a detailed study of DLA in presence of an
   external magnetic field, we solved the corresponding Langevin equation exactly. It is found that the
   solution involves  correlated Gaussian random variables. In order to carry out numerical simulation
   of DLA, it is necessary to sample these random variables, at discrete instants of time,
   consistent with the correlation relations. In this paper, we describe an exact sampling procedure
   that is valid for arbitrary values of all the parameters that appear in the problem. The results of the
   DLA simulation using this prescription are in qualitative agreement with the experimental results
    and will be published elsewhere \cite{mini}.

\noindent
\section{Solution of Langevin equation in the presence of magnetic field}

Consider the Langevin equation describing the Brownian motion of a charged particle of unit mass and
charge q, in presence of a magnetic field $\vec{B}$ along the z-direction
\begin{eqnarray}
{d\over{dt}}\vec{r}(t) & = & \vec{v}(t) \\
{d\over{dt}}\vec{v}(t) +\gamma \vec{v}(t) - q \vec{v}(t)\times \vec{B} & = & \vec{\eta}(t)
\end{eqnarray}
\noindent
In the above equation, $ \vec{\eta}(t) = \left(\eta_1(t), \eta_2(t), \eta_3(t)\right) $ is a
Gaussian white noise of zero mean and its components satisfy the normalization condition

\begin{eqnarray}
\left \langle\eta _i(t)\eta _j(t')\right \rangle = 2A \delta _{ij} \delta (t-t')
\end{eqnarray}

\noindent
Since the magnetic field is oriented along the z-direction, it affects the motion in the x
and y directions only. Defining
\begin{eqnarray}
X = x + i y, \ \ V = v_1 + i v_2, \ \  \Gamma =\gamma + i \omega, \ \ \eta = \eta_1 + i \eta_2
\end{eqnarray}
we get
\begin{eqnarray}
{d\over{dt}}X & = & V\\
{d\over{dt}}V + \Gamma V & = & \eta(t)
\end{eqnarray}
The formal solution is given by
\begin{equation}
V(t)  =  e^{-\Gamma t} V(0) + \int_0^t e^{-\Gamma (t-t_1)} \eta (t_1) \ dt_1
\end{equation}
\begin{widetext}
\begin{equation}
X(t) - X(0)  =  {{1}\over{\Gamma}}\left( (1-e^{-\Gamma t}) V(0) +\left[
\int_0^t{\eta (t_1) \ dt_1} -\int_0^t {e^{-\Gamma (t-t_1)} \eta(t_1) \ dt_1}\right] \right)
\end{equation}
\end{widetext}

\noindent
Using the above formal solution, we can immediately obtain $x(t), y(t)$ by taking the real and
imaginary parts of $X(t)$. $v_1(t)$ and $v_2(t)$ can also be obtained in a similar fashion.
The two time correlation functions of $v_1(t)$ and $v_2(t)$ are given by
\begin{eqnarray}
\left \langle v_1(t) v_1(t') \right \rangle & = & \left(\frac{A}{\gamma}\right)e^{-\gamma \vert{t-t'}\vert} \cos(\omega (t-t'))\\
\left \langle v_2(t) v_2(t') \right \rangle  & = & \left \langle v_1(t) v_1(t') \right \rangle\\
\left \langle v_1(t) v_2(t') \right \rangle & = &\left(\frac{A}{\gamma}\right) e^{-\gamma \vert{t-t'}\vert} \sin(\omega (t-t'))
\end{eqnarray}
The other correlation functions can also be calculated easily.\\

\noindent
The interesting point is that unlike in the case of the motion  without the magnetic field, now the
components $v_1(t)$ and $v_2(t')$, of the velocity, are correlated. But, the second moment
$<v_i(t)^2>$, of each of the components of the velocity, has the same value $A/\gamma$ as in the
case of zero magnetic field. This can be easily understood once we recognize that  $<v_i(t)^2>$  is
twice the kinetic energy  and that does not change by application of a magnetic field. However,
$<x(t)^2>$ and $<y(t)^2>$ asymptotically (i.e., as $t \longrightarrow \infty$) tend to
${{2 A}\over{\vert\Gamma\vert^2}}t ={{2 A}\over{(\gamma^2+\omega^2)}}t$ so that the diffusion coefficient
$D={{A}\over{(\gamma^2+\omega^2)}}$ decreases with increasing magnetic field.

\noindent
In the numerical simulations we will be interested in getting realizations of $x(t)$, $y(t)$, etc.
at discrete instants of time $t_j = j*\tau$.  We have developed such a scheme that is valid for
arbitrary values of $A$, $\omega$, $\gamma$ and $\tau$.

\noindent
Define
\begin{eqnarray}
x_j = x(t_j), \ \   y_j  &  = &  y(t_j),  \ \    X_j=x_j + i y_j\\
\psi_j & = &   \int_0^\tau{du \  \eta(j\tau+u)}\\
\phi_j& = &  \int_0^\tau{du \  e^{-\Gamma(\tau-u)} \eta(j\tau+u)}\\
\psi_j^{(t)}& = &  \sum_{l=0}^{j-1}{\psi_l}\\
\phi_j^{(t)}& = &  \sum_{l=0}^{j-1}{e^{-(j-1-l)\Gamma\tau}\phi_l}
\end{eqnarray}
We can then write
\begin{eqnarray}
X_j = {{1}\over{\Gamma}}\left[(1 - e^{-j\Gamma\tau}) V(0)+\psi_j^{(t)}- \phi_j^{(t)}\right]
\end{eqnarray}
We can perform the numerical simulations if we can reliably sample $\psi_j$ and $\phi_j$. since
$\psi_j$ and  $\phi_j$  depend on $\eta$ in the interval $[j\tau, (j+1)]\tau]$ alone, and since the
$\lbrace\eta_i\rbrace$ are delta-correlated, we can immediately see that  $\left \langle\psi_j\psi_j'\right \rangle$ and
$\left \langle\phi_j\phi_j'\right \rangle$ vanish when $j \ne j'$. However, $\psi_j$ and $\phi_j$ are correlated.
Therefore it is important to evolve a procedure to sample these correlated random variables.

\noindent
It is convenient to work with the following real random variates:

\begin{eqnarray}
\psi_{jR} = \Re({\psi_j}) & = & \int_0^\tau{du \  \eta_1(j\tau+u)}\\
\psi_{jI} = \Im({\psi_j}) & = & \int_0^\tau{du \  \eta_2(j\tau+u)}
\end{eqnarray}
\begin{widetext}
\begin{eqnarray}
\phi_{jR} = \Re({\phi_j}) & = & \int_0^\tau{du \  e^{-\gamma(\tau-u)}
\left[\cos(\omega(\tau-u))\eta_1(j\tau+u) + \sin(\omega(\tau-u))\eta_2(j\tau+u)\right]}\\
\phi_{jI} = \Im({\phi_j}) & = & \int_0^\tau{du \  e^{-\gamma(\tau-u)}
\left[-\sin(\omega(\tau-u))\eta_1(j\tau+u) + \cos(\omega(\tau-u))\eta_2(j\tau+u)\right]}
\end{eqnarray}

[In the above equations, $\Re(z)$ and $\Im(z)$ are the real and imaginary parts of the complex
number $z$]. Since   $\psi_j$ and  $\phi_j$  are Gaussian random variates of zero mean, the
knowledge of their covariance is adequate for describing them. They are given below:

\begin{eqnarray}
\label{corr1}
\left \langle\psi_{jR} \psi_{j'R}\right \rangle & = & 2A\tau\delta_{j,j'}, \ \
\left \langle\psi_{jI} \psi_{j'I}\right \rangle  =  2A\tau\delta_{j,j'}, \ \
\left \langle\psi_{jR} \psi_{j'I}\right \rangle  =  0\\
\label{corr2}
\left \langle\phi_{jR} \phi_{j'R}\right \rangle & = & {{A}\over{\gamma}}(1 - e^{-2\gamma\tau})\delta_{j,j'}, \ \
\left \langle\phi_{jI} \phi_{j'I}\right \rangle  =  {{A}\over{\gamma}}(1 - e^{-2\gamma\tau})\delta_{j,j'}, \ \
\left \langle\phi_{jR} \phi_{j'I}\right \rangle  =  0\\
\label{corr3}
\left \langle\phi_{jR} \psi_{j'R}\right \rangle & = & {{2A}\over{(\gamma^2 + \omega^2)}}\left[\gamma(1 -
e^{-\gamma\tau} \cos(\omega\tau)) + \omega e^{-\gamma\tau} \sin(\omega\tau)\right]\delta_{j,j'}\\
\label{corr4}
\left \langle\phi_{jR} \psi_{j'I}\right \rangle & = & {{2A}\over{(\gamma^2 + \omega^2)}}\left[\omega(1 -
e^{-\gamma\tau} \cos(\omega\tau)) - \gamma e^{-\gamma\tau} \sin(\omega\tau)\right]\delta_{j,j'}\\
\label{corr5}
\left \langle\phi_{jI} \psi_{j'R}\right \rangle  & = & - \left \langle\phi_{jR} \psi_{j'I}\right \rangle, \ \
\left \langle\phi_{jI} \psi_{j'I}\right \rangle   =  \left \langle\phi_{jR} \psi_{j'R}\right \rangle
\end{eqnarray}
\end{widetext}

\noindent
Our aim is to provide a scheme for sampling  $\psi_{jR}$, $\psi_{jI}$,  $\phi_{jR}$ and
$\phi_{jI}$ that is consistent with the above equations. We propose to achieve this by
expressing $\psi_{jR}$, $\psi_{jI}$,  $\phi_{jR}$  and $\phi_{jI}$ as sums of  n
identically distributed Gaussian random variables $(p_{j1}, p_{j2}, ..., p_{jn})$ with zero mean
and unit variance in the following manner:
\begin{eqnarray}
\psi_{jR} & = & \sqrt{2A\tau}\sum_{l=1}^n {\alpha_l p_{jl}}\\
\psi_{jI} & = & \sqrt{2A\tau}\sum_{l=1}^n {\beta_l p_{jl}}\\
\phi_{jR} & = & \sqrt\frac{A(1-e^{-2\gamma\tau})}{\gamma}\sum_{l=1}^n {\epsilon_l p_{jl}}\\
\phi_{jI} & = & \sqrt\frac{A(1-e^{-2\gamma\tau})}{\gamma}\sum_{l=1}^n {\nu_l p_{jl}}\\
\end{eqnarray}
The substitution of the above results in equations \ref{corr1} - \ref{corr5} gives 10 constraints that the parameters
 $\alpha_l$, $\beta_l$, $\epsilon_l$ and $\nu_l$ should satisfy.
 \begin{widetext}
\begin{eqnarray}
\label{alphabeta}
\sum_{l=1}^n{\alpha_l^2} & = & 1, \ \
\sum_{l=1}^n{\beta_l^2}  =  1, \ \
\sum_{l=1}^n{\alpha_l\beta_l}  =  0\\
\label{epsilonnu}
\sum_{l=1}^n{\epsilon_l^2} & = & 1, \ \
\sum_{l=1}^n{\nu_l^2}  =  1, \ \
\sum_{l=1}^n{\epsilon_l\nu_l}  =  0\\
\label{epsilonalpha}
\sum_{l=1}^n{\epsilon_l\alpha_l} & = & \sqrt\frac{2\gamma}{\tau(1-e^{-2\gamma\tau})}\frac{\left[\gamma(1 -
e^{-\gamma\tau} \cos(\omega\tau)) + \omega e^{-\gamma\tau} \sin(\omega\tau)\right]}{(\gamma^2+\omega^2)} \equiv C_1\\
\label{epsilonbeta}
\sum_{l=1}^n{\epsilon_l\beta_l} & = &\sqrt\frac{2\gamma}{\tau(1-e^{-2\gamma\tau})}\frac{\left[\omega(1 -
e^{-\gamma\tau} \cos(\omega\tau)) - \gamma e^{-\gamma\tau} \sin(\omega\tau)\right]}{(\gamma^2+\omega^2)} \equiv C_2\\
\label{nubeta}
\sum_{l=1}^n{\nu_l\alpha_l}  & = & -\sum_{l=1}^n{\epsilon_l\beta_l} = -C_2, \ \
\sum_{l=1}^n{\nu_l\beta_l}   = \sum_{l=1}^n{\epsilon_l\alpha_l} = C_1
\end{eqnarray}
\end{widetext}
$\lbrace\alpha_i\rbrace$, $\lbrace\beta_i\rbrace$, $\lbrace\epsilon_i\rbrace$  and $\lbrace\nu_i\rbrace$ may be
considered as unimodular vectors $\vec\alpha$, $\vec\beta$, $\vec\epsilon$ and $\vec\nu$  in an n-dimensional
 space. Such a geometrical interpretation of the above relations immediately leads to the conclusion that 4
is the minimum value of n, for which the above relations can be satisfied, see the Appendix.
We consider the case n=4, henceforth. Since  there are 16 parameters and only 10 constraints,
there will be a 6-parameter family of solutions. We can choose  $\vec\alpha$ and $\vec\beta$ consistent
only with the constraints represented by eq. \ref{alphabeta}, and then solve for
  $\vec\epsilon$ and $\vec\nu$. In what follows, we show two solutions explicitly and also
  discuss the prescription to obtain any other solution.\\

\subsection{Solution 1}
A simple choice for $\vec\alpha$ and $\vec\beta$ would be
\begin{eqnarray}
\alpha_1 = 1, \   \alpha_2 = 0, \  \alpha_3 = 0, \  \alpha_4 = 0,   \\
\beta_1 = 0, \  \beta_2 = 1,  \  \beta_3 = 0,  \ \beta_4=0
\end{eqnarray}
This choice leads to
\begin{eqnarray}
\epsilon_1 = C_1,   \epsilon_2 = C_2,   \epsilon_3 = Rcos(\theta),   \epsilon_4 = Rsin(\theta), \\
\nu_1 = -C_2,  \nu_2 = C_1,   \nu_3 = -Rsin(\theta),   \nu_4 = Rcos(\theta),
\end{eqnarray}
where $R=\sqrt{(1-C_1^2-C_2^2)}$ and $\theta$ is a free parameter.\\

\subsection{Solution 2}
Another choice for $\vec\alpha$ and $\vec\beta$ is
\begin{eqnarray}
\alpha_1   =     \alpha_2  =    \alpha_3 & = &   \alpha_4   = \frac{1}{2}\\
\beta_1   =     -\beta_2   =     -\beta_3 & = &   \beta_4   = \frac{1}{2}
\end{eqnarray}
We then obtained  $\vec\epsilon$ and $\vec\nu$ to be
\begin{eqnarray}
\epsilon_1  =  D_1 + D_3 \cos(\theta), & \ \  & \epsilon_2  =  D_2 + D_3 \sin(\theta),\\
\epsilon_3   =   D_2 - D_3 \sin(\theta), & \ \ &  \epsilon_4  =  D_1- D_3 \cos(\theta)\\
\nu_1   =   D_2 + D_3 \sin(\theta), & \ \ & \nu_2   =  -D_1 - D_3 \cos(\theta),\\
\nu_3   =  -D_1 + D_3 \cos(\theta), & \ \ & \nu_4   =   D_2 + D_3 \sin(\theta),
\end{eqnarray}
where $D_1$  =  $(C_1 + C_2)/2$,   $D_2  =  (C_1 -C_2)/2$, $D_3   =   \sqrt{1-(C_1^2+C_2^2)/2}$
and $\theta$ is a free parameter which was set to be $\pi/4$ in our simulations.\\

 Any set of  vectors $\vec\alpha'$, $\vec\beta'$, $\vec\epsilon'$ and $\vec\nu'$ obtained by simultaneously
  rotating (SO(4) rotation) a given set of vectors $\vec\alpha$, $\vec\beta$, $\vec\epsilon$ and $\vec\nu$ will also be a
  valid solution of the equations \ref{alphabeta} - \ref{nubeta}. It remains to be seen if a particular choice of
  these vectors is better than others for numerical simulations with finite number of realizations or not.

\section{Conclusions}

The Langevin equation for diffusion of charged particles in presence of an external magnetic field is
 solved exactly. A scheme for sampling the position and velocity is given which is valid for arbitrary
 values of the  friction coefficient of the medium, the strength of the magnetic field and the time step of
 evolution. The components of the position and velocity are represented as linear combinations of four
 independent and identically distributed Gaussian random variates. A six parameter family of solutions
  is obtained for the expansion coefficients. Further studies are required to see if a particular choice of the
  expansion coefficients is better than all the other possibilities.

\section{Appendix}
Consider four unimodular  vectors $\vec\alpha$, $\vec\beta$, $\vec\epsilon$,
 $\vec\nu$ satisfying the conditions
\begin{eqnarray}
\label{A1}
\vec\alpha.\vec\alpha=1, \ \ \vec\beta.\vec\beta=1, \ \ \vec\alpha.\vec\beta=0\\
\label{A2}
\vec\epsilon.\vec\epsilon=1, \ \ \vec\nu.\vec\nu=1, \ \ \vec\epsilon.\vec\nu=0\\
\label{A3}
\vec\epsilon.\vec\alpha=C_1, \ \ \vec\epsilon.\vec\beta=C_2\\
\label{A4}
\vec\nu.\vec\alpha=-C_2, \ \ \vec\nu.\vec\beta=C_1
\end{eqnarray}
\subsection{2-dimensional case}
Let us first consider the 2-dimensional case. Here, the most general  pair of  vectors $\vec\alpha$
 and $\vec\beta$ satisfying eq. \ref{A1} can be easily constructed as follows
\begin{eqnarray}
\label{A5}
\vec\alpha=(cos\theta_1, sin\theta_1),\ \  \vec\beta= (-sin\theta_1  ,cos\theta_1).
\end{eqnarray}
where $\theta_1$ is an arbitrary parameter. Since $\vec\alpha$
 and $\vec\beta$ are orthonormal, they form a basis in 2-dimensions and hence
 any other vector can be expressed as linear combinations of these vectors.
Therefore $\vec\epsilon$  and  $\vec\nu$ can be written as
\begin{eqnarray}
\label{A6}
\vec\epsilon=a_1 \vec\alpha + a_2 \vec\beta,\ \  \vec\nu = b_1 \vec\alpha + b_2 \vec\beta
\end{eqnarray}
The expansion coefficients $a_1$, $a_2$, $b_1$ and $b_2$ can be obtained by substituting
eq. \ref{A6} in eqs. \ref{A3} and \ref{A4}. The result is
\begin{eqnarray}
\label{A7}
\vec\epsilon= C_1 \vec\alpha + C_2 \vec\beta,\ \ \vec\nu = -C_2 \vec\alpha + C_1 \vec\beta
\end{eqnarray}
Eq. \ref{A7} implies
\begin{eqnarray}
\label{A8}
\vec\epsilon .\vec\epsilon= \vec\nu .\vec\nu = C_1^2 + C_2^2
\end{eqnarray}
which is not equal to 1 in general. Thus the $\vec\epsilon$  and  $\vec\nu$ fail to satisfy the
 unimodularity condition (see eq. \ref{A2}). Hence we see that it is impossible to obtain the
 four vectors $\vec\alpha$, $\vec\beta$, $\vec\epsilon$
and $\vec\nu$ satisfying the constraints eqs. \ref{A1}-\ref{A4} in two dimension.\\
\subsection{3-dimensional case}

The most general vector $\vec\alpha$ can be written down as
\begin{equation}
\label{A9}
\vec\alpha=(sin\theta_1cos\phi_1, sin\theta_1sin\phi_1, cos\theta_1)
\end{equation}
where $\theta_1$ and $\phi_1$ are arbitrary parameters. The most general vector $\vec\beta$
perpendicular to $\vec\alpha$ is given by
\begin{equation}
\label{A10}
\vec\beta=cos(\zeta) \vec\beta + sin(\zeta) (\vec\alpha\times \vec\eta )
\end{equation}
with
\begin{equation}
\label{A11}
 \vec\eta=(cos\theta_1cos\phi_1, cos\theta_1sin\phi_1, -sin\theta_1).
\end{equation}
and $\zeta$ is a free parameter. It is obvious that the vectors
$\vec\alpha$, $\vec\beta$, $\vec\alpha \times \vec\beta$ form a basis in three dimensional space.
Then $\vec\epsilon$ can be defined as
\begin{equation}
\label{A12}
\vec\epsilon = a_1 \vec\alpha + a_2 \vec\beta + a_3 \vec\alpha \times \vec\beta.
\end{equation}
Using eqs. \ref{A3} and then \ref{A2}, we get
\begin{equation}
\label{A13}
\vec\epsilon=C_1\vec\alpha+C_2\vec\beta+a_3\vec\alpha\times \vec\beta.
\end{equation}
with $C_1^2 + C_2^2 + a_3^2$ =1. Similar analysis gives
\begin{equation}
\label{A14}
\vec\nu=-C_2\vec\alpha+C_1\vec\beta+b_3\vec\alpha\times \vec\beta
\end{equation}
with $C_1^2 + C_2^2 + b_3^2$ =1. Substituting for  $\vec\epsilon$ and $\vec\nu$, we get
\begin{equation}
\vec\epsilon.\vec\nu=-C_1C_2 + C_1C_2 + a_3b_3 = a_3b_3
\end{equation}
which is not zero in general. Hence  it is impossible to obtain the
 four vectors $\vec\alpha$, $\vec\beta$, $\vec\epsilon$
and $\vec\nu$ satisfying the constraints eqs. \ref{A1}-\ref{A4} in three dimensions as well.
It is, however, possible to obtain solutions in four dimensions, as explicitly demonstrated in the text.
\end{document}